\documentclass[11point,a4paper]{article}
\usepackage{jheppub}
\usepackage{amsmath} 
\usepackage{amsfonts} 
\usepackage{amssymb}
\usepackage{bbm}
\usepackage{epsfig}
\usepackage{graphics}
\usepackage{graphicx}
\usepackage{hyperref}

\newcommand{\lsim}{\,\lower2truept\hbox{${<\atop\hbox{\raise4truept\hbox{$\sim$}}}$}\,}
\newcommand{\gsim}{\,\lower2truept\hbox{${>\atop\hbox{\raise4truept\hbox{$\sim$}}}$}\,}
\newcommand{\pp}{~~~.}
\newcommand{\vv}{~~~,}

\newcommand{\be}{\begin{equation}}
\newcommand{\ee}{\end{equation}}
\newcommand{\bea}{\begin{eqnarray}}
\newcommand{\eea}{\end{eqnarray}}
\newcommand{\beann}{\begin{eqnarray*}}
\newcommand{\eeann}{\end{eqnarray*}}
\newcommand{\nn}{\nonumber}

\newcommand{\mn}{{\mu\nu}}

\newcommand{\rs}{{\rho\sigma}}

\newcommand{\hmn}{h_{\mu\nu}}
\newcommand{\hmnup}{h^{\mu\nu}}
\newcommand{\hmnt}{\tilde{h}_{\mu\nu}}
\newcommand{\hmnupt}{\tilde{h}^{\mu\nu}}

\newcommand{\Ltri}{\mathcal{L}^{(3)}}
\newcommand{\humn}{h^{\mu\nu}}
\newcommand{\p}{\partial}
\newcommand{\half}{\frac{1}{2}}
\newcommand{\huab}{h^{\alpha\beta}}

\title{On ghosts in theories of self-interacting massive spin-2 particles}

\author{Sarah Folkerts, Alexander Pritzel and Nico Wintergerst}
\affiliation{Arnold-Sommerfeld-Center, Ludwig-Maximilians-Universit\"at, Theresienstr. 37, 80333 M\"unchen, Germany}

\emailAdd{sarah.folkerts@physik.lmu.de}
\emailAdd{alexander.pritzel@physik.lmu.de}
\emailAdd{nico.wintergerst@physik.lmu.de}

\abstract{
We consider general theories of a massive spin-2 particle $\hmn$ on a Minkowski background.
A decomposition of $\hmn$ in terms of helicity eigenstates allows us to directly test whether any given theory possesses a consistent description as a massive spin-2 representation of the Poincar\'e group.
We demonstrate (i) that any nonlinear theory with an Einsteinian derivative structure either contains ghosts or does not describe a weakly coupled spin-2 and (ii) that there exists a two-parameter family of non-Einsteinian cubic self-interactions which constitute a ghost-free massive spin-2 theory.
}

\subheader{LMU-ASC 30/11}

\begin{document}

\maketitle
\flushbottom

\section{Introduction} 

The search for viable theories of massive gravity has proven to be very difficult. 
Massive gravity as the theory of an interaction mediated by a massive spin-2 field must necessarily propagate five degrees of freedom (DOF), namely its five polarizations. While this fixes the mass term uniquely on the linear level to the Fierz-Pauli form \cite{fierz_pauli_1939}, it also introduces immediate troubles. When taking the massless limit of the theory, the additional graviton polarizations do not generically decouple. As a consequence, predictions of any massive gravity theory in a first linear approximation differ from massless general relativity (GR) by numerical factors \cite{vandam_veltman_zakharov_1970}. However, it has been pointed out in \cite{vainshtein_1972} that this so-called van Dam-Veltman-Zakharov (vDVZ) discontinuity may disappear when correctly taking nonlinearities into account, with nonlinearities growing with decreasing graviton mass as $m^{-4}$. Later, in \cite{deffayet_etal_2001} this nonperturbatively continuous behavior has been demonstrated for a specific model of massive gravity.

The question of instabilities in nonlinear theories of massive gravity has been addressed in many works. Boulware and Deser \cite{boulware_deser_1973} proved the inevitable appearance of the sixth polarization of the graviton as a ghost-like state in a wide class of models through a Hamiltonian formalism. On the level of the action, the appearance of a six-derivative cubic term in the helicity-0 component in Einstein gravity with a Fierz-Pauli mass term has been shown in \cite{deffayet_etal_2001} in terms of the leading singularity of the graviton vertex, and in terms of St\"uckelberg fields in \cite{arkanihamed_etal_2002}. Both in \cite{arkanihamed_etal_2002, creminelli_etal_2005} it was proven that one may cancel such operators by appropriately adding non-derivative interactions to the action. However, it remained unsettled if other operators could spoil the stability of the theory. Since then, several other works have tried to construct manifestly stable theories of a single massive graviton (\cite{derham_gabadadze_etal,Chamseddine:2010ub} and references therein). Recently, \cite{derham_etal_2010} suggested that one may formulate nonlinear theories containing the Fierz-Pauli mass term, which, on the full nonlinear level, were found to describe the correct number of degrees of freedom in \cite{HassanRosen}. Others have argued that generic theories of massive gravity contain problems such as ghosts or superluminality \cite{alberte_etal_2010,gruzinov_2011}. While the latter is beyond the scope of this work, we will confirm findings on the ghost problem in massive gravity.

We will demonstrate that an expansion of nonlinear Fierz-Pauli models in terms of a weakly coupled massive spin-2 field inevitably leads to inconsistencies.
We consider theories which are Poincar\'{e} invariant, local and weakly coupled in at least some energy interval where they describe a single propagating massive spin-2 particle, $\hmn$, on a Minkowski background. They can be expressed in terms of polynomial interactions of a local Lagrangian.  
For a stability analysis, it is important to realize that ghost-type instabilities are relevant on arbitrarily short timescales: While tachyonic instabilities arise for momenta lower than the tachyonic mass $m_\text{tach}$ and contain an intrinsic timescale $t = 1/m_\text{tach}$, ghosts are UV instabilities. Their intrinsic timescale is $t = 1/E$, which is only limited by an effective field theory (EFT) cutoff $\Lambda$. It is therefore justified to test the system for instabilities at momenta $m \ll E \ll \Lambda$.
This gives a straightforward prescription for a stability analysis in terms of representations of the Poincar\'{e} group. At very high energies, $E \gg m$, the massive spin-2 representation of the Poincar\'e group decomposes into the direct sum of irreducible helicity representations.
In other words, the ratio of mass and energy $m/E$ parametrizes the mixing between helicity eigenstates. For sufficiently large $E$, this mixing becomes negligible, helicity eigenstates decouple on a linear level and constitute a unique description of the system.
As we only consider weakly coupled theories, we can capture all relevant physics by decomposing $\hmn$ \emph{linearly} into its helicities. Any inconsistencies found in these states inevitably indicate (i) ghost-like instabilities or (ii) a violation of the assumptions of Poincar\'e invariance, locality and weak coupling. Either way, a consistent description in terms of a weakly coupled massive spin-2 field $\hmn$ is excluded. In this sense, decomposing the massive spin-2 into its helicities has a big advantage compared to previous methods. It enables one to not only count degrees of freedom, but gives a direct test whether they can be grouped into a massive spin-2 particle.

This work will therefore elucidate aspects of the ongoing debate by explicitly showing that no nonlinear extension of Fierz-Pauli theory with an Einsteinian derivative structure is free of inconsistencies. Two interpretations are possible. The first is that the theory contains ghosts, which must be addressed in any analysis of viability. One has to introduce additional degrees of freedom that cure the ultraviolet (UV) regime, or ensure that the ghost degrees of freedom are otherwise shielded. Note here that we are not interested in the position of ghost poles, but are simply looking for the mere appearance of pathological degrees of freedom.
The second interpretation, possibly applicable to \cite{HassanRosen}, is that the weakly coupled degrees of freedom cannot be attributed to a massive spin-2 Poincar\'e representation or the theory exhibits strong coupling already in the infrared.

However, an Einsteinian derivative structure is not preferred when considering general theories of massive spin-2 fields. For $m=0$ it is known that Poincar\'{e} invariance, locality and unitarity alone pin down general relativity as the unique theory of self-interactions \cite{weinberg_1964_1965, Deser}. For $m \neq 0$ these arguments cannot be generalized.
Henceforth, we will construct a ghost-free cubic theory of a massive spin-2 field by only requiring the self-interactions to be Lorentz-invariant and to involve at most two derivatives. A priori, all coupling parameters are arbitrary and can be adjusted model dependently. 
We will exploit this freedom in parameters to eliminate possible ghost-like instabilities and extra DOFs and prove that the constructed action is a valid theory of a weakly coupled massive spin-2 particle.

Our paper is organized as follows. In section \ref{sec:FP}, we review the theory of a free massive spin-2 particle due to Fierz and Pauli \cite{fierz_pauli_1939} and explain how it can be understood in terms of helicity degrees of freedom, whose mixing is governed by the mass of the particle. We will further explain why a decomposition into helicities as kinetic eigenstates is necessarily linear.

We apply the helicity decomposition to the class of theories whose derivative interactions are governed by Einsteinian vertices in section \ref{sec:ECI}. We demonstrate that inconsistencies inevitably appear on the cubic level, even when allowing
for additional arbitrary nonderivative interactions. Furthermore we discuss recent claims \cite{derham_gabadadze_etal, HassanRosen} in the literature regarding the eligibility of such a theory.

Finally, in section \ref{sec:GCI}, we construct a two-parameter family of cubic interactions which manifestly describe the ghost-free theory of a self-interacting massive spin-2 particle. This two-parameter family does not contain the cubic part of the Einstein-Hilbert action and 
is therefore not a likely candidate for a massive version of gravity.

\section{Fierz-Pauli} \label{sec:FP}

\subsection{Action}
The following set of conditions must hold for any theory subject to this work:
\begin{enumerate}
  \renewcommand{\theenumi}{[\alph{enumi}]}
\item \label{mass_s2} It must be a Poincar\'e invariant theory of a single massive spin-2 field on a flat Minkowiski background. That is, the field lives in the irreducible Poincar\'e representation labeled by the Casimir operators mass $m \neq 0$ and spin $s = 2$.
\item \label{local} It must be local, that is it can be understood in terms of polynomial interactions of the spin-2 field.
\item \label{weak_coup} It must be weakly coupled in at least a finite momentum range $m^2 \ll k^2 \ll \Lambda^2$, with an EFT cutoff $\Lambda$. In other words, ${\cal L}_n \ll {\cal L}_{n-1}$, where ${\cal L}_n$ describes the $n$-th power interactions of $\hmn$. 
Note here that from now on energies and momenta are to be understood to lie beneath the cutoff.
\end{enumerate}
In particular, \ref{weak_coup} allows us to limit the analysis to cubic self-interactions, as the possibility of removing a ghost by addition of higher order terms is explicitly excluded. As will be seen, this is sufficient to rule out an Einsteinian derivative structure for the tri-$h$-vertex.

We denote the massive spin-2 particle by a symmetric rank-2 tensor $\hmn$. The free part of any viable action is uniquely fixed to take on the form \cite{fierz_pauli_1939}
\be
\label{fierz-pauli}
{\cal L}_\text{PF} = \hmnup {\cal E}^{\rs}_{\mn} h_\rs - \frac{1}{2} m^2 \left( \hmnup\hmn - h^2 \right) \vv
\ee
where ${\cal E}^{\rs}_{\mn}$ is the linearized Einstein operator, defined by 
\be
\label{einstein_op}
{\cal E}^{\rs}_{\mn}h_\rs \equiv \half \Box \hmn - \p_\alpha \p_{(\mu} h^\alpha_{\nu)}+\p_\mu\p_\nu h-\half \eta_\mn \Box h \pp
\ee
The equation of motion for $\hmn$ becomes 
\be
{\cal E}^{\rs}_{\mn}h_\rs - m^2 (\hmn - h \eta_{\mu\nu}) = 0 \pp
\ee
By virtue of the Bianchi identities, acting upon the equation of motion with the operators $\p_\mu$ and $\frac{1}{2} \eta_\mn + \frac{\p_\mu\p_\nu}{m^2}$, respectively, yields five constraint equations:
\bea
\p_\mu h^\mu_\nu &=& \p_\nu h \vv \\
\label{eq:tr_const} h &=& 0 \pp
\eea
Hence five of the ten components of $\hmn$ are propagating. If departing from the Fierz-Pauli mass term, eq.(\ref{eq:tr_const}) changes to become an equation of motion for the trace $h$, turning it into a propagating degree of freedom. Depending on the relative factor between $\hmnup\hmn$ and $h^2$, the sixth degree of freedom $h$ is either tachyonic or ghost-like.

Note another way of counting the degrees of freedom encoded in eq.(\ref{fierz-pauli}): In terms of components of $\hmn$, the exact structure (\ref{fierz-pauli}) uniquely ensures that $h_{00}$ appears linearly in the action, while $h_{0i}$ appears without time derivatives. One hence has four non-propagating components of which one acts as a Lagrange multiplier, reducing the number of degrees of freedom to five. Departing from the Fierz-Pauli mass term introduces nonlinearities in $h_{00}$. One loses the constraint which fixed the spatial trace $h_{ii}$ to zero, again resulting in a tachyonic or ghost-like sixth degree of freedom.

\subsection{Helicities}

As discussed above, counting degrees of freedom in terms of $\hmn$ requires an analysis of constraint equations. This, while in principle possible, can become quite cumbersome when taking interactions into account. There is, however, a formalism which allows us to straightforwardly see if a given theory is a valid theory of a weakly coupled massive spin-2 particle.

For high momenta, the irreducible massive spin-2 representation of the Poincar\'e group effectively decomposes into its helicity-2, helicity-1 and helicity-0 parts. In other words, for $k^2 \gg m^2$ the massive representation may be reduced to a direct sum of the massless helicity-representations. All relevant physics is encoded in helicity eigenstates. In particular, short time-scale instabilities, i.e. ghosts, can be seen directly on the helicities. We stress that the combination of these properties gives us an immensely powerful tool to analyze any theory of massive gravity described as a rank-2 tensor. It is not only sensitive to the correct number of degrees of freedom, but also to the question whether these degrees of freedom are grouped into a massive spin-2. 
In summary, inconsistencies seen in the theory in terms of helicities can have the following origins:
\begin{enumerate}
  \renewcommand{\theenumi}{[\roman{enumi}]}
\item \label{first_crit} The theory contains ghosts.
\item There is no weak coupling regime for $k^2 \gg m^2$, i.e. \ref{weak_coup} is violated.
\item The weakly coupled degrees of freedom cannot be grouped to form a massive spin-2 particle (cf.\ref{mass_s2}). This happens explicitly for example in Lorentz violating theories.
\item Additional degrees of freedom are required to enter the theory at some scale (cf.\ref{mass_s2}) or the theory is shielded otherwise. 
\item \label{final_crit} The theory is nonlocal (cf.\ref{local}).
\end{enumerate}
As we will elaborate on in Section \ref{sec:ECI}, this enables us to show that the nonlinear theory of \cite{HassanRosen}, while possessing the correct number of degrees of freedom, cannot describe a weakly coupled massive spin-2 field. It falls into at least one of the above categories.

To change into the helicity basis, one decomposes $\hmn$ into a sum of another tensor $\hmnt$, a vector $A_\mu$ and a scalar $\chi$. These fields describe the correct helicities when coefficients are adjusted s.t. the kinetic term diagonalizes. In that case, the mixing of the different polarizations of $\hmn$ manifests itself solely through terms proportional to $m$. For high momenta, these mixing terms become irrelevant and the helicities become independent degrees of freedom.

Let us elaborate on the reasons for the decomposition to be linear in helicities.
First, the helicity basis is defined via the requirement that the kinetic operator is diagonal; helicities are kinetic eigenstates. However, the kinetic operator of $\hmn$ will only be diagonal in terms of helicities if there is a linear relation.
Second, we can understand any field theory in terms of propagators and vertices of the considered fields, in this case $\hmn$.  A decomposition which allows direct probing of instabilities in the relevant degrees of freedom should not interfere with this statement; the propagator and vertices of $\hmn$ should directly correspond to the propagators and vertices of the helicities. This allows only for linear decompositions, as can be seen very clearly when taking the coupling to external sources into account. The same argument holds for asymptotic states in scattering experiments. Observation or probing of $\hmn$ at high energies should be equivalent to observation of individual helicities; 
an asymptotic state of $\hmn$'s must not differ from an asymptotic state prepared in terms of helicities. This is only the case if the helicities are contained linearly in $\hmn$.
Finally, we emphasize that the requirement of weak coupling forbids nonlinearities in the fields $A$ and $\chi$ in the decomposition. This would introduce ambiguities between different orders of $\hmn$ which are not consistent with a weak field approximation. 
There will be further comments on this point in Section \ref{sec:discussion}.

The decomposition takes on the following form \cite{dvali_2006}:
\be
\label{decomp} \hmn = \hmnt + \frac{\p_{(\mu}A_{\nu)}}{m} + \frac{1}{3} \left(\frac{\partial_\mu \partial_\nu \chi}{m^2} + \frac{1}{2} \eta_\mn \chi \right) \vv
\ee
where $\hmnt$ describes the helicity-2, $A_\mu$ the helicity-1 and $\chi$ the helicity-0 part of the massive spin-2 Poincar\'{e} representation. 
As discussed above, the power of the decomposition \eqref{decomp} can be seen explicitly when inserted into the quadratic action \eqref{fierz-pauli},
\bea
{\cal L}_\text{PF} &=& \hmnupt {\cal E}^{\rs}_{\mn} \tilde{h}_\rs - \frac{1}{8} F_\mn F^\mn + \frac{1}{12}\chi\Box\chi - \frac{1}{2} m^2 \left( \hmnupt\hmnt - \tilde{h}^2 \right)  + \frac{1}{6}m^2\chi^2 \nn \\
&& + \frac{1}{2}m^2\chi\tilde{h} + m\left(\tilde{h}\partial_\mu A^\mu - \hmnupt\partial_\mu A_\nu\right) + \frac{m}{2} \chi \partial_\mu A^\mu \pp
\eea
For $k^2 \gg m^2$, the action diagonalizes. The individual kinetic terms for $\hmnt$ and $A_\mu$ correspond to massless linearized Einstein and Maxwell theory, respectively. Thus, in this limit, $\hmnt$ carries precisely the two helicity-2 DOFs, $A_\mu$ the two helicity-1 DOFs and $\chi$ the single helicity-0 component. 

Note that requiring the diagonalization of the kinetic term fixes the relative factor of $1/2$ between the $\chi$-terms in \eqref{decomp}. Similarly, the factors of $m$ in \eqref{decomp} normalize the kinetic terms. The coefficient of the kinetic term for $\chi$ is determined by the coupling of $\hmn$ to sources: $\int d^4x T^{\mu\nu}\hmn$. The propagator of $\hmn$ between two conserved sources $T_{\mu\nu}$ and $\tau_{\mu\nu}$ is given by
\bea
T^{\mu\nu}D_{\mu\nu,\rho\sigma}\tau^{\rho\sigma} &=& T^{\mu\nu}\frac{\left(\eta_{\mu\rho}\eta_{\nu\sigma}+\eta_{\mu\sigma}\eta_{\nu\rho}-\frac{2}{3}\eta_{\mu\nu}\eta_{\rho\sigma}\right)}{p^2-m^2}\tau^{\rho\sigma} \nonumber\\
&=&T^{\mu\nu}\frac{\left(\eta_{\mu\rho}\eta_{\nu\sigma}+\eta_{\mu\sigma}\eta_{\nu\rho}-\frac{1}{2}\eta_{\mu\nu}\eta_{\rho\sigma}\right)}{p^2-m^2}\tau^{\rho\sigma} + T^{\mu\nu}\frac{1}{6}\frac{\eta_{\mu\nu}\eta_{\rho\sigma}}{p^2-m^2}\tau^{\rho\sigma}\; .
\eea
The first term corresponds to the helicity-2 state. The second term is an additional interaction from the extra scalar DOF and fixes the overall normalization of $\chi$ in our helicity decomposition. By considering non-conserved sources one can accordingly fix the normalization of $A_\mu$ in \eqref{decomp}.

As a side remark, note that $\hmn$ is invariant under the following set of simultaneous gauge transformations:
\bea
\hmnt &\rightarrow& \hmnt + \partial_{(\mu}\xi_{\nu)} + \frac{1}{2}\eta_\mn m \Lambda \vv\nonumber \\
A_\mu &\rightarrow& A_\mu + \partial_\mu\Lambda - m \xi_\mu \vv\nonumber\\
\label{gauge_trafos}\chi &\rightarrow& \chi - 3 m \Lambda \pp
\eea
These redundancies are expected, as only ten out of 15 components on the right hand side of eq.(\ref{decomp}) are fixed. Both sides will in the end describe a maximum of ten degrees of freedom. Also note here again that by construction the validity of the decomposition is limited to a theory of a weakly coupled massive spin-2 particle. If $\hmn$ is used to describe different degrees of freedom, eq.(\ref{decomp}) is no longer guaranteed to capture the correct physics. This is consistent with the group theoretical arguments outlined above.

\section{Einsteinian interactions} \label{sec:ECI}

\subsection{Cubic Vertex and Ghosts}
As a first application of our method we consider the question of the appearance of a sixth graviton polarization, commonly referred to as the Boulware-Deser ghost \cite{boulware_deser_1973}, in nonlinear extensions of massive gravity. In this context it is assumed that the derivative structure of the nonlinear theory must correspond to an expansion of the Ricci scalar. One further allows for an addition of arbitrary nonderivative self-interactions, leading to an action
\be
\label{nonlin_act} S = 2 M_P^2 \int d^4x \sqrt{-g} R + S_{\text{nonder}}[\hmn] \vv
\ee
with $g_{\mu\nu} = \eta_{\mu\nu} + h_{\mu\nu}$ and $S_{\text{nonder}}[\hmn]$ contains all nonderivative interactions.
Boulware and Deser argued that the appearance of an additional sixth polarization as a nonlinear ghost mode is inevitable. However, recent works \cite{derham_gabadadze_etal, HassanRosen} claim to have found a flaw in the original argument which allows for an extinction of the ghost under certain conditions. We will elaborate on this in Section \ref{sec:discussion}.

In this section, we will show that if conditions \ref{mass_s2}-\ref{weak_coup} are to be fulfilled by the theory, inconsistencies cannot be avoided.

As discussed above, we have two advantages at hand which greatly simplify the arguments. Due to the assumption of weak coupling, it is sufficient to only consider cubic interactions. Further, we may work in a helicity basis. At high energies,
or equivalently short time scales (relevant for the investigation of ghost instabilities), the spin-2 field $\hmn$ decomposes into its helicity components, $\hmnt$, $A_\mu$ and $\chi$. These states also couple to external sources and can thus be excited on the linear level. 
Any inconsistencies found in terms of these fields inevitably satisfy one of the criteria \ref{first_crit}-\ref{final_crit}.
We will see that for the cubic order Einsteinian theory $A_\mu$ and $\chi$ always appear with higher derivatives in their equation of motion, either indicating ghost-like instabilities on general backgrounds or signaling a departure from a valid massive spin-2 description of the theory. 

Expanding eq.\eqref{nonlin_act} to cubic order yields the interaction Lagrangian 
\bea
\label{cubic Einstein}
\Ltri&=&\frac{1}{M_P} \big[\{\frac{1}{4}\huab\p_\alpha\humn\p_\beta\hmn- \frac{1}{4}\huab\p_\alpha h\p_\beta h + \huab\p_\beta h \p_\mu h_\alpha^\mu-\frac{1}{2} \humn \p_\alpha h\p^\alpha\hmn \nn \\
&& + \frac{1}{8}h \p_\mu h \p^\mu h - \humn\p_\alpha h^\alpha_\mu \p_\beta h_\nu^\beta-\humn\p_\nu h_\mu^\alpha \p_\beta h_\alpha^\beta +\frac{1}{2} h \p_\mu \humn\p_\alpha h_\nu^\alpha \nn \\
&& + \half\humn \p^\alpha\hmn\p_\beta h_\alpha^\beta - \frac{1}{4}h\p_\alpha h \p_\beta \huab+\half \humn \p_\alpha h_{\nu\beta}\p^\beta h_\mu^\alpha + \half\humn \p_\beta h_{\nu\alpha}\p^\beta h_\mu^\alpha \nn \\
&& -\frac{1}{4}h\p_\alpha \hmn \p^\nu h^{\mu\alpha} - \frac{1}{8}h\p_\alpha \humn \p^\alpha\hmn\} +m^2(k_{1} h_\nu^\mu h^\nu_\rho h^\rho_\mu +k_{2}  h h_{\mn}\humn+k_{3}h^3)\big]
\; ,
\eea
where $k_1, k_2, k_3$ are free parameters.
Inserting the decomposition \eqref{decomp},
one immediately encounters higher derivative terms in the $\chi$ and $A_\mu$ sectors. 
Operators appearing with seven and eight derivatives are boundary terms and can be disregarded.
The terms with the highest derivative contribution to the equations of motion (EOM) are cubic in $\chi$, i.e. self-interactions of the helicity-0 modes, and are suppressed by the scale $\Lambda_5^5 \equiv m^4 M_P$. This is the lowest scale in the theory and constitutes the EFT cutoff. These interactions are the first ones to become strong at high energies and are thus the most important ones for the stability analysis. By taking the limit $M_P \rightarrow \infty$ and $m\rightarrow 0$ while keeping $\Lambda_5$ fixed, the so-called decoupling limit, one can focus only on the $\chi$ self-interactions at that scale. 
The resulting Lagrangian takes the form
\bea\label{L5dec}
\mathcal{L}_\text{dec} &=& \mathcal{L}_\text{kin}(\hmnt,A_\mu,\chi) + \frac{1}{432\Lambda _5^5} \bigg[(2+8 k_1+16 k_2+32 k_3)(\Box \chi)^3 \nonumber\\
&&+(2-24 k_1-16 k_2)\chi \Box\chi \Box^2\chi +(1-12 k_1-8 k_2)\chi^2\Box^3 \chi \bigg] \vv
\eea
where $\mathcal{L}_\text{kin}$ contains the kinetic terms of all helicities.

One can use the freedom in the parameters $k_1, k_2, k_3$ to eliminate the higher derivatives on the EOM of $\chi$. The required relations are
\bea
\label{1st dec rule}
1+4k_1+8k_2+16k_3&=&0 \nonumber\\
1-12 k_1-8 k_2&=&0\; .
\eea
Consequently, 
all interactions suppressed by the scale $\Lambda_5$ vanish. 
Subsequently, the lowest scale is $\Lambda_4^4 \equiv m^3 M_P$. We again focus on the leading remaining interactions by taking another decoupling limit: $m \rightarrow 0$, $M_P \rightarrow \infty$ and $\Lambda_4$ fixed. The Lagrangian is given by
\be
\label{decL4}
\mathcal{L}_\text{dec4} = \mathcal{L}_\text{kin}\left(\hmnt,A_\mu,\chi\right) + \frac{1}{36 \Lambda_4^4}\left[A^\mu\p_\mu\p^\nu\chi\Box\p_\nu\chi+\frac{1}{2}\p_\mu A^\mu\p_\rho  \p_\nu \chi\p^\rho  \p^\nu \chi\right]  \; .
\ee
The higher derivative interactions present on the EOM do not depend on the remaining free parameter $k_1$. One must conclude that the interactions given by the Lagrangian \eqref{cubic Einstein} lead to instabilities in the helicity components of the theory and should be discarded as a possible theory for
a weakly coupled spin-2 particle.
While the free theory on Minkowski is perfectly fine and understandable in terms of irreducible representations of the Poincar\'e group, 
adding interactions appears to induce additional DOFs. This confirms also the results of \cite{alberte_etal_2010}.

\subsection{Raising the Cutoff}

Our method reveals a further peculiarity of the theory \eqref{nonlin_act}. Previous works \cite{arkanihamed_etal_2002, derham_gabadadze_etal} were seemingly able to completely remove the strong coupling scale $\Lambda_5$ from the action, leaving a theory with an apparent EFT cutoff at $\Lambda_3$. In terms of helicities, however, we can see that any theory of the form \eqref{nonlin_act} will still contain the scale $\Lambda_5$.

Note first that on the quartic level, a generic Einsteinian operator $\p^2 h^4/M_P^2$ will contain maximum derivative contributions $\p^{10} \chi^4$, which are suppressed by the scale $\Lambda_5^{10}$ (cf. eq.\eqref{decomp}). However, a nonderivative term $h^4$ can at best produce $\p^8 \chi^4$ operators, which cannot cancel ten-derivative contributions; $\Lambda_5$ will inevitably appear in quartic interactions.

Explicit computation of the expansion of $\sqrt{-g}R$ confirms the presence of nontrivial operators suppressed by $\Lambda_5$.
After eliminating cubic terms by choosing coefficients according to \eqref{1st dec rule}, corresponding to those found in \cite{derham_gabadadze_etal}, 
the full decoupling limit Lagrangian of a general theory \eqref{nonlin_act} reads
\bea\label{L5dec4}
\mathcal{L}_\text{dec} &=& \mathcal{L}_\text{kin}(\hmnt,A_\mu,\chi) +\frac{1}{2}\frac{1}{6^4 \Lambda_5^{10}}\chi \bigg[\partial ^\nu\Box\chi \big\{\partial_\nu(\partial ^\sigma\Box\chi)^2 - 2\partial_\nu( \partial _\lambda\partial _\sigma\partial _\rho\chi)^2 -2 \partial ^\sigma\partial ^\rho\partial _\nu\chi  \Box\partial _\sigma\partial _\rho\chi \big\}\nonumber\\
&&+4 \partial _\lambda\partial _\sigma\partial _\rho\partial _\nu\chi \partial ^\rho\partial ^\nu\partial_\mu\chi  \partial ^\mu\partial ^\lambda\partial ^\sigma\chi -(\partial _\rho\partial _\nu\partial _\mu\chi)^2 \Box^2\chi +(\p_\nu\Box\chi)^2  \Box^2\chi \bigg]\; .
\eea
There is no freedom to eliminate the quartic terms. Any action \eqref{nonlin_act} of a weakly coupled massive spin-2 field, will, as understood in helicities, contain the scale $\Lambda_5$.

\subsection{Discussion} \label{sec:discussion}

This section provides a discussion of the results found above, in particular in context with previous works.
Let us first comment on the analysis of \cite{derham_gabadadze_etal}. 
It is not straightforward to compare the two results. An action constructed in terms of St\"uckelberg tensors $H_{\mu\nu}(\hmn, A_\mu, \phi)$ and $H^{\mu\nu} = g^{\mu\rho} g^{\nu\sigma} H_{\rho\sigma}$, will, even to the second order in $H_{\mu\nu}$, contain infinite powers of $\hmn$. Therefore there is no simple relation of the degrees of freedom. However, we may be able to elucidate some of the aspects by considering a St\"uckelberg trick of the traditional form
\bea
\hmn &=& \hmnt + \frac{1}{m^2} \p_{(\mu}\xi_{\nu)} + \frac{1}{\Lambda_5^5} \p_\mu\xi_\rho\p_\nu\xi^\rho + f\left(\hmnt,\xi_\mu\right) \nonumber \\
\label{eq:stueck} \xi_\mu &=& m A_\mu + \p_\mu\phi \vv
\eea
where $f$ contains all mixing between $\hmnt$ and $\xi_\mu$. Note that $\xi_\mu$ is to transform s.t. $\hmn$ is invariant under general coordinate transformations on $\hmnt$.
Inserting \eqref{eq:stueck} into the action \eqref{cubic Einstein} with coefficients \eqref{1st dec rule} removes the strong coupling scale $\Lambda_5$ from the action and renders the Lagrangian \eqref{L5dec4} into that of a free field. In this sense $\phi$ can be compared to the scalar mode of \cite{derham_gabadadze_etal}. One might think that this gives us a straightforward way to relate the fields $\chi$ and $\phi$ by a nonlinear field redefinition. Equating \eqref{decomp} and \eqref{eq:stueck} indeed yields the relation
\be
\label{fieldredef}
\eta_\mn \chi = \eta_\mn \phi+\frac{2}{\Lambda_5^5}\left(\p_{\mu}\p_\sigma\phi\p_\nu\p^\sigma\phi-\frac{\p_\rho\p_{(\mu}}{\Box+m^2/2}\p_{\nu)}\p_\sigma\phi\p^\rho\p^\sigma\phi\right) \pp
\ee
One can define a redefinition by taking the trace of \eqref{fieldredef}. 
Unfortunately, this relation contains several puzzles. Its traceless part yields a set of conditions on $\phi$. The resulting theory, albeit seemingly that of a free field, has a corresponding space of solutions which is limited by these conditions and in fact contains interactions.
Furthermore, the form of \eqref{fieldredef} suggests that it is not invertible.

Regardless of these issues, 
there is another problem with the nonlinear decomposition \eqref{eq:stueck}.
A theory describing a weakly coupled massive spin-2 field 
should have a well-defined perturbative expansion in terms of $\hmn$.
Terms of order $\hmn^4$ are by definition less important than terms of order $\hmn^3$ and hence cannot be used to cure ghost instabilities at third order. Otherwise the requirement of weak coupling of all DOFs is violated by at least one of the DOFs.
However, a nonlinear decomposition invalidates this requirement.
Terms of second order in $h_{\mu\nu}$ will introduce terms of up to fourth order in the helicity-1 and -0 fields which can then be canceled by helicity operators introduced in the third and fourth order in $\hmn$.  This appears to be the case in \cite{arkanihamed_etal_2002} when raising the effective field theory cutoff to $\Lambda_3$, as well as in \cite{derham_gabadadze_etal} when eliminating higher derivative interactions. 

The approach of \cite{derham_gabadadze_etal} was extended to the full nonlinear level in \cite{derham_etal_2010}, where a family of potentially ghost-free nonlinear extensions of the Fierz-Pauli mass term was suggested.
Without loss of generality, we consider a subclass further studied in \cite{HassanRosen}. For a flat auxiliary metric the Lagrangian is given by
\be
\label{Hassan}
\mathcal{L}=-M_P^2\int d^4x\sqrt{-g}\left[R+2m^2 Tr\sqrt{g^{-1}\eta}-6m^2\right]\pp
\ee
When expanded around Minkowski, \eqref{Hassan} reproduces the decoupling limit for the scalar mode as considered in \cite{derham_gabadadze_etal}.
Following a procedure proposed in \cite{derham_etal_2010}, the authors of \cite{HassanRosen} employed an ADM decomposition \cite{ADM} of the dynamical metric $g_{\mu\nu}$ to find that on the full non-linear level the theory describes five dynamical DOFs. The crucial difference to the analysis of Boulware and Deser \cite{boulware_deser_1973} is that DOFs are explicitly counted on the constraint surface. After integrating out the shift $N_i$, the lapse $N$ is again found to be a Lagrange multiplier.
While this procedure appears to be legitimate for the setup considered in \cite{HassanRosen}, it might run into trouble when considering coupling to matter\footnote{We thank L. Alberte for pointing this out.}. Choosing the coupling equivalent to general relativity does not increase the number of degrees of freedom. However, it seems that such a choice is not protected and by EFT reasoning other operators should be included. 

There are further issues that have yet to be addressed in an analysis of \eqref{Hassan}. Expanding the action in terms of a weakly coupled massive spin-2 field $\hmn$ reveals, as shown above, inconsistencies. Either, there seems to be no weak coupling regime $m^2 \ll k^2 \ll \Lambda^2$, or the degrees of freedom cannot be understood as a massive spin-2 representation of Poincar\'e. Understanding this will require further investigation. 

Finally, we point to another unusual property of \eqref{Hassan}. It contains an auxiliary Minkowski metric without dynamics. While this might be unproblematic in terms of a field theory on a fixed Minkowski background, it cannot be straightforwardly generalized to nonflat backgrounds. Whether this problem is related to the points above is an interesting question.

\section{General Cubic Interactions}
\label{sec:GCI}

We have now understood that an Einsteinian derivative structure for the cubic theory results in inconsistent interactions of helicities. One can wonder whether relaxing restrictions on the derivative interactions can lead to an interacting cubic order theory of $\hmn$ which does not encounter this problem.
We will construct such a Lagrangian in this section. To our knowledge, this is the first time such an action with derivative interactions has been presented in the literature. 
Our starting point consists of the most general cubic interaction Lagrangian which is Lorentz-invariant and includes at most two derivatives. 
It reads
\be
\label{startL}
\mathcal{L} = \mathcal{L}_\text{FP}+\mathcal{L}^{(3)} \vv
\ee
where ${\cal L}_\text{FP}$ is defined according to \eqref{fierz-pauli} and, up to boundary terms, 
\bea
\label{L3}
\Ltri&=&k_1\huab\p_\alpha\humn\p_\beta\hmn+k_2 \huab\p_\alpha h\p_\beta h +k_3\huab\p_\beta h \p_\mu h_\alpha^\mu+k_4 \humn \p_\alpha h\p^\alpha\hmn+ k_5h \p_\mu h \p^\mu h\nonumber\\
&&+k_6\humn\p_\alpha h^\alpha_\mu \p_\beta h_\nu^\beta+k_7\humn\p_\nu h_\mu^\alpha \p_\beta h_\alpha^\beta+k_8 h \p_\mu \humn\p_\alpha h_\nu^\alpha+ k_9\humn \p^\alpha\hmn\p_\beta h_\alpha^\beta\nonumber\\
&&+k_{10}h\p_\alpha h \p_\beta \huab+k_{11} \humn \p_\alpha h_{\nu\beta}\p^\beta h_\mu^\alpha+k_{12}\humn \p_\beta h_{\nu\alpha}\p^\beta h_\mu^\alpha+k_{13}h\p_\alpha \hmn \p^\nu h^{\mu\alpha}\nonumber\\
&&+k_{14}h\p_\alpha \humn \p^\alpha\hmn+ k_{15} h_\nu^\mu h^\nu_\rho h^\rho_\mu+k_{16}  h h_{\mn}\humn+k_{17}h^3 \; .
\eea
The coefficients $k_i$ are free parameters which will be adjusted in such a way that the interacting theory remains ghost-free and propagates only five degrees of freedom. 
To prevent the appearance of additional degrees of freedom in the theory, one limits the allowed number of time derivatives acting on helicities in the equations of motions to two.
We proceed as follows: By inserting the decomposition \eqref{decomp} into the Lagrangian \eqref{startL}, we derive the EOM for the individual helicity eigenstates, $A_\mu$ and $\chi$, and check for higher time derivatives on the fields. Then, by exploiting the freedom in parameter space we try to gradually eliminate these terms. The advantage of working directly on the EOM is 
that all higher derivative terms appearing are relevant, as boundary terms do not contribute.
The details of our calculation can be found in the Appendix. The constraint of allowing only for at most two time-derivatives on the EOM determines all coefficients in terms of $k_1$ and $k_{15}$ and gives the following cubic Lagrangian
\bea
\label{lastcoe}
\Ltri&=& k_1\big(\huab\p_\alpha\humn\p_\beta\hmn- \huab\p_\alpha h\p_\beta h +4\huab\p_\beta h \p_\mu h_\alpha^\mu-2 \humn \p_\alpha h\p^\alpha\hmn+h \p_\mu h \p^\mu h\nonumber\\
&&-3\humn\p_\alpha h^\alpha_\mu \p_\beta h_\nu^\beta-4\humn\p_\nu h_\mu^\alpha \p_\beta h_\alpha^\beta +3 h \p_\mu \humn\p_\alpha h_\nu^\alpha+ 2\humn \p^\alpha\hmn\p_\beta h_\alpha^\beta\nonumber\\
&&-2h\p_\alpha h \p_\beta \huab+ \humn \p_\alpha h_{\nu\beta}\p^\beta h_\mu^\alpha+2\humn \p_\beta h_{\nu\alpha}\p^\beta h_\mu^\alpha-h\p_\alpha \hmn \p^\nu h^{\mu\alpha}-h\p_\alpha \humn \p^\alpha\hmn\big) \nn \\
&&+ \half k_{15}\big(2 h_\nu^\mu h^\nu_\rho h^\rho_\mu-3  h h_{\mn}\humn+h^3\big) \pp
\eea
That this theory still propagates the right number of degrees of freedom can also be easily seen by counting the number of constraints for $h_{\mu\nu}$. 
As explained in section \ref{sec:FP}, in Fierz-Pauli five
 constraints for $\hmn$ reduce the number of DOFs to five. 
 For the Lagrangian \eqref{lastcoe}, these constraints are preserved:  $h_{0i}$ is still non-dynamical and can be solved for algebraically, yielding 3 constraints on $\hmn$. Furthermore, $h_{00}$ still appears as a Lagrange multiplier in (\ref{lastcoe})
 and accordingly eliminates another two DOFs. 
 
Eq.\eqref{lastcoe} is to be understood as an effective action with a cutoff given by the lower of $m(m/k_{15})^{1/3}$ and $m (m k_1)^{-1/3}$. In order for the theory to be a useful description, the cutoff should be larger than $m$, implying the following hierarchy for the parameters involved: $k_{15}\ll m \ll k_1^{-1}$.

Of course we have only proven the absence of ghosts. As any cubic theory, \eqref{lastcoe} will still contain tachyonic instabilities. One may however easily extend our formalism to higher orders.

\section{Conclusions}

The aim of this work was to investigate the consistency of nonlinear extensions of Fierz-Pauli theory on a Minkowski background. In contrast to previous works \cite{arkanihamed_etal_2002, derham_gabadadze_etal}, no attention was paid to restoration of general covariance. Instead we made direct use of the isometries of the flat background and worked in terms of irreducible representations of the Poincar\'e group. In addition, the only requirements were locality and weak coupling, s.t. the theory could be understood as polynomial interactions of the field $\hmn$. Any theory meeting these requirements is inevitably subject to our results.

Our set of prerequisites allowed us to greatly reduce the complexity of the analysis. A weakly coupled theory can, for sufficiently weak fields, be understood in terms of its lowest order interaction, which enabled us to focus solely on the cubic action. Further, we made use of the fact that for high momenta, the massive spin-2 representation decomposes into a direct sum of helicity-2, -1 and -0 representations and all physical information is contained in helicity eigenstates. Finally, weak coupling, amongst other things, requires the decomposition to be linear, as this excludes mixing between different order interactions.

We first applied this formalism to a massive gravity theory with derivative interactions governed by the expanded Ricci scalar. Despite allowing the addition of nonderivative interactions, we found higher derivative operators already on the cubic level. Furthermore, we demonstrated that in the same class of theories, the scale $\Lambda_5^5 \equiv m^4 M_P$ cannot be fully eliminated from the action when taking quartic interactions into account.

Our result only allows for two conclusions. Either the theory violates the prerequisites, i.e. the DOFs are not a massive spin-2 particle, they are not weakly coupled or the theory is nonlocal, or there are additional ghost degrees of freedom present. The former interpretation in particular applies to the class of models introduced in \cite{HassanRosen}, since the authors were able to prove the absence of an additional sixth degree of freedom on the full nonlinear action. It is an interesting question what these theories really describe. First hints may be drawn from the fact that a nondynamical auxiliary metric is necessarily introduced into the action. Allowing for dynamics of this auxiliary field, albeit introducing additional degrees of freedom, may in the end provide valuable insight into this class of models.

Finally, we applied our formalism to the most general cubic Lagrangian for the field $\hmn$. We showed that one can tune all interactions in such a way that a manifestly ghost-free two-parameter family of theories of a massive spin-2 field is found. This could be seen both in terms of helicities and in terms of components of $\hmn$. A Dirac constraint analysis revealed the same set of conditions as in linear Fierz-Pauli theory. A possible phenomenological application of this action has to be analyzed in more detail. Since the derivative structure differs from the Einsteinian cubic vertex, it is highly doubtful that it can be applied to the problem of giving a mass to the graviton. Furthermore, it still encounters the problem of tachyonic instabilities, as any cubic theory. One may however easily extend our analysis to higher order interactions and this way be able construct manifestly stable theories of a self-interacting massive spin-2 field.

\acknowledgments
We wish to thank Gia Dvali for proposing the underlying formalism and for many helpful discussions. We further thank Lasma Alberte, Felix Berkhahn, Cristiano Germani, Stefan Hofmann, Parvin Moyassari and Rachel Rosen for useful comments and stimulating discussions. Tensorial computations have been carried out with the help of the \textsc{Mathematica} package \texttt{xAct} \cite{xTensor}. The work of all authors is supported by the Alexander von Humboldt foundation.

\appendix
\section{Appendix}
Within this appendix, we present the computation leading to the Lagrangian \eqref{lastcoe}.
Starting from the interaction Lagrangian \eqref{L3}, we first derive the equations of motion for the helicity-0 component $\chi$ and subsequently eliminate higher time derivatives.
Eradicating $\Box^2\chi\Box^2 \chi$, $\Box \chi \Box^3 \chi$, $\p_\mu\Box\chi \partial^\mu\Box^2 \chi$ and $\Box \chi \Box^2\chi$ fixes four coefficients:
\bea
\label{Rule1} 2 k_{10}-k_2-k_3+k_4+2 k_5-k_6+k_7+2 k_8+k_9&=&0 \vv  \\ 
\label{Rule2} k_{13}+k_{14}+k_5+k_8+\frac{1}{2} \left(k_2+k_3-k_4-2 k_5+k_6-k_7-2 k_8-k_9\right)&=& 0\vv \\ 
\label{Rule3}k_1+k_{11}+k_{12}+k_2+k_3+k_4+k_6+k_7+k_9&=&0  \vv \\
\label{Rule4} 8 k_{16}+24 k_{17}+\left(8 k_1-k_2-7 k_3+3 k_4-18 k_5-13 k_6+17 k_7+18 k_8+9 k_9\right) m^2&=&0\pp 
\eea
We proceed with eliminating terms such as $\p_\mu\Box\chi\p^\mu\Box\chi$, $\chi\Box^2\chi$
\bea
\label{Rule5}6 k_{15}+4 k_{16}-\left(2 k_1+11 k_2+5 k_3+3 k_4-k_6+2 k_7\right) m^2&=&0  \vv\\
\label{Rule6}-2 k_1+k_3+2 k_5+2 k_6-3 k_7-2 k_8-2 k_9&=&0 \pp
\eea
Next, we consider the EOM for the vector $A^\mu$. Eliminating $\partial_\mu A^\mu\Box^2A_\alpha$, $\partial_\alpha A^\mu\Box^2A_\mu$, $\Box A_\alpha\Box\p_\mu A^\mu$, $\Box A^\mu\Box\p_\alpha A_\mu$ and $\Box A_\mu\Box\p^\mu A_\alpha$ sets five coefficients:
\bea
\label{Rule7}
2 k_1+2 k_{13}+2 k_2+k_3-2 k_4-2 k_5+k_7&=&0 \vv  \\
\label{Rule8}
2 k_2+3 k_3+2 k_4+2 k_5+2 k_6-k_7-2 k_8&=&0 \vv\\
\label{Rule9}
2 k_1-2 k_4-2 k_5+k_7&=&0  \vv \\
\label{Rule10}
2 k_2+k_3+2 \left(k_4+k_5\right)&=&0 \vv  \\
\label{Rule11}
k_{11}+2 k_2+k_3+k_6&=&0 \pp
\eea
Reverting to mixed interactions, the EOM for $\chi$ contains terms such as $\Box\tilde{h}\Box^2\chi$, $\Box^2\tilde{h}\Box\chi$, $\p^\mu\p^\nu\chi\Box^2 \tilde{h}_{\mu\nu}$ and $\p_\mu\chi \p^\mu\Box\tilde{h}$ requiring
\bea
\label{Rule12}
-k_1+k_{11}+4 k_2+k_3&=&0 \vv\\
\label{Rule13} 4 k_2+2 k_3+2 k_4&=& 0 \vv \\
\label{Rule14}k_1+k_2&=& 0\vv \\
\label{Rule15}4 k_1-k_3&=&0 \pp
\eea
This leaves the Lagrangian \eqref{lastcoe}, whose corresponding equations of motion are free of higher time derivatives.


\begin{thebibliography}{2}

\bibitem{fierz_pauli_1939}
  M.~Fierz, W.~Pauli,
  ``On relativistic wave equations for particles of arbitrary spin in an electromagnetic field,''
  Proc.\ Roy.\ Soc.\ Lond.\  {\bf A173 } (1939)  211-232.

\bibitem{vandam_veltman_zakharov_1970}
  H.~van Dam and M.~J.~G.~Veltman,
  ``Massive And Massless Yang-Mills And Gravitational Fields,''
  Nucl.\ Phys.\  B {\bf 22} (1970) 397.
  V.~I.~Zakharov,
  ``Linearized gravitation theory and the graviton mass,''
  JETP Lett.\  {\bf 12} (1970) 312
  [Pisma Zh.\ Eksp.\ Teor.\ Fiz.\  {\bf 12} (1970) 447].
  

\bibitem{vainshtein_1972}
  A.~I.~Vainshtein,
  ``To the problem of nonvanishing gravitation mass,''
  Phys.\ Lett.\  B {\bf 39} (1972) 393.

\bibitem{deffayet_etal_2001}
  C.~Deffayet, G.~R.~Dvali, G.~Gabadadze and A.~I.~Vainshtein,
  ``Nonperturbative continuity in graviton mass versus perturbative
  discontinuity,''
  Phys.\ Rev.\  D {\bf 65} (2002) 044026
  [arXiv:hep-th/0106001].

\bibitem{boulware_deser_1973}
  D.~G.~Boulware and S.~Deser,
  ``Can gravitation have a finite range?,''
  Phys.\ Rev.\  D {\bf 6}, 3368 (1972).

\bibitem{arkanihamed_etal_2002}
  N.~Arkani-Hamed, H.~Georgi, M.~D.~Schwartz,
  ``Effective field theory for massive gravitons and gravity in theory space,''
  Annals Phys.\  {\bf 305 } (2003)  96-118.
  [hep-th/0210184].

\bibitem{creminelli_etal_2005}
  P.~Creminelli, A.~Nicolis, M.~Papucci and E.~Trincherini,
  ``Ghosts in massive gravity,''
  JHEP {\bf 0509} (2005) 003
  [arXiv:hep-th/0505147].

\bibitem{derham_gabadadze_etal}
  C.~de Rham, G.~Gabadadze,
  ``Generalization of the Fierz-Pauli Action,''
  Phys.\ Rev.\  {\bf D82 } (2010)  044020.
  [arXiv:1007.0443 [hep-th]].

\bibitem{Chamseddine:2010ub}
  A.~H.~Chamseddine, V.~Mukhanov,
  ``Higgs for Graviton: Simple and Elegant Solution,''
  JHEP {\bf 1008 } (2010)  011.
  [arXiv:1002.3877 [hep-th]].

  \bibitem{derham_etal_2010}
 C.~de Rham, G.~Gabadadze, A.~J.~Tolley,
  ``Resummation of Massive Gravity,''
  Phys.\ Rev.\ Lett.\  {\bf 106 } (2011)  231101.
  [arXiv:1011.1232 [hep-th]].

\bibitem{HassanRosen}
  S.~F.~Hassan, R.~A.~Rosen,
  ``On Non-Linear Actions for Massive Gravity,''
   [arXiv:1103.6055 [hep-th]].
  S.~F.~Hassan, R.~A.~Rosen,
  ``Resolving the Ghost Problem in non-Linear Massive Gravity,''
   [arXiv:1106.3344 [hep-th]].

\bibitem{gruzinov_2011}
  A.~Gruzinov,
  ``All Fierz-Paulian massive gravity theories have ghosts or superluminal modes,''
  [arXiv:1106.3972 [hep-th]].

\bibitem{alberte_etal_2010}
 L.~Alberte, A.~H.~Chamseddine, V.~Mukhanov,
  ``Massive Gravity: Exorcising the Ghost,''
  JHEP {\bf 1104 } (2011)  004.
  [arXiv:1011.0183 [hep-th]].
  A.~H.~Chamseddine and V.~Mukhanov,
  ``Massive Gravity Simplified: A Quadratic Action,''
  arXiv:1106.5868 [hep-th].
 

\bibitem{weinberg_1964_1965}
  S.~Weinberg,
  ``Photons And Gravitons In S Matrix Theory: Derivation Of Charge Conservation
  And Equality Of Gravitational And Inertial Mass,''
  Phys.\ Rev.\  {\bf 135} (1964) B1049.
  S.~Weinberg,
  ``Photons and gravitons in perturbation theory: Derivation of Maxwell's and
  Einstein's equations,''
  Phys.\ Rev.\  {\bf 138} (1965) B988.

  \bibitem{Deser}
 S.~Deser,
 ``Self-interaction and gauge invariance,''
 Gen.\ Rel.\ Grav.\  {\bf 1}, 9 (1970)
 [arXiv:gr-qc/0411023].

\bibitem{dvali_2006}
  G.~Dvali,
  ``Predictive Power of Strong Coupling in Theories with Large Distance
  Modified Gravity,''
  New J.\ Phys.\  {\bf 8}, 326 (2006)
  [arXiv:hep-th/0610013].
  

\bibitem{ADM}
 R.~L.~Arnowitt, S.~Deser, C.~W.~Misner,
  ``Dynamical Structure and Definition of Energy in General Relativity,''
  Phys.\ Rev.\  {\bf 116 } (1959)  1322-1330.

 \bibitem{xTensor}
J.-M.~Mart\'{\i}n-Garc\'{\i}a,
Comp. Phys. Commun. {\bf 179}, 597 (2008)
[arXiv:0803.0862 [cs.SC]],
$<$http://metric.iem.csic.es/Martin-Garcia/xAct/$>$.
  

\end{thebibliography}
\end{document}